\documentclass[letterpaper]{article}
\usepackage{aaai}
\usepackage{times}
\usepackage{helvet}
\usepackage{courier}
\usepackage{subfig}
\usepackage{graphicx}
\begin{document}
%
\title{\emph{Blind Men and the Elephant}: Detecting Evolving Groups in Social News}
\author{
Roja Bandari\\
Dept. of Electrical Engineering\\ University of California, Los Angeles \\ Los Angeles, CA 90095-1594\\roja@ucla.edu
\And Hazhir Rahmandad\\
Dept. of Industrial and Systems Engineering\\Virginia Polytechnic Institute\\ Falls Church, VA 22043 \\ hazhir@vt.edu
\And Vwani P. Roychowdhury\\ 
Dept. of Electrical Engineering\\University of California, Los Angeles\\ Los Angeles, CA 90095-1594\\vwani@ee.ucla.edu
}
%

\maketitle
\begin{abstract}
\begin{quote} 
We propose an automated and unsupervised methodology for a novel summarization of group behavior based on content preference. We show that graph theoretical community evolution (based on similarity of user preference for content) is effective in indexing these dynamics. Combined with text analysis that targets automatically-identified representative content for each community, our method produces a novel multi-layered representation of evolving group behavior. We demonstrate this methodology in the context of political discourse on a social news site with data that spans more than four years and find coexisting political leanings over extended periods and a disruptive external event that lead to a significant reorganization of existing patterns. Finally, where there exists no ground truth, we propose a new evaluation approach by using entropy measures as evidence of coherence along the evolution path of these groups. This methodology is valuable to designers and managers of online forums in need of granular analytics of user activity, as well as to researchers in social and political sciences who wish to extend their inquiries to large-scale data available on the web.  
\end{quote}
\end{abstract}

\section{Introduction}
\noindent  

Online forums and social news sites have created new spaces for user interaction that can influence millions of individuals as well as traditional media platforms. The importance of these spaces is evident in the surge of web-data analysis throughout the 2012 US presidential election \cite{ScienceElections2012}\footnote{Also see http://cacm.acm.org/blogs/blog-cacm/156624-priming-assimilation-bias-social-proof-in-social-media}. These spaces of discussion and information sharing provide large datasets that can be investigated by scholars in various fields. While important questions have been formed and extensively studied by social scientists in smaller scales, any such attempt on the web is met with the computational challenges of processing and abstracting very large, complex, and often noisy data, rendering methods developed for smaller scales impractical.  


While an array of techniques have been developed for generating a variety of summary statistics from large-scale data, they all fall short of offering a complete multi-layered summary that enables scholarly investigation in social sciences or allows the owner of a site to understand her user base in terms of how they interact with content and with each other, and how such interaction patterns evolve over time. Consider a website with many users who share articles online and express their opinions in various ways. Other than the unscalable approach of manually and laboriously following almost all the activities of the user population and becoming experts on the related topics, how would one begin to understand its dynamics? One can begin by reporting simple statistical measures (most popular articles, most active or most influential users, percentage of items containing some keyword, increase or decrease in activity), employ language processing to measure positive or negative sentiment, detect topics of discussion, or use regression to model or predict specific measures.  Like the parable of \emph{Blind Men and the Elephant}\footnote{ http://en.wikipedia.org/wiki/Blind\_men\_and\_an\_elephant}, these techniques provide us with disjoint, specific pieces of information. We believe there is a need for development of {\it automated tools}  that are not manually  coded with domain-specific knowledge (hence, applicable to sites across several verticals), and yet provide a top-down summary of user dynamics. Such an automated multi-scale summary then can facilitate a more granular exploration. To the best of our knowledge, such a framework has not been offered by the scientific community.


We use explicit indicators of user preference for content as the basis for our methodology. Some examples of such indicators are the ``Like" button in Facebook, an ``up" vote in reddit, a ``+1" in Google-plus, or a ``digg" on Digg. We will call these indicators \emph{votes} in the context of this paper and will use them as clear and simple signals that can be used to infer user orientation toward content. For example, intuition suggests that users who prefer and promote the same political articles will have similar political leanings, whereas explicit friendships do not necessarily suggest similar political orientations. 


Based on votes cast by users in a bipartite network of users and articles, we detect communities of users with similar voting patterns and track these communities' temporal evolution. We then identify representative content for each community based on their votes, and perform more detailed analysis on text and source of these representative sets, teasing out persistent themes\footnote{We call the groups of users \emph{communities} because they are produced through community detection methods. These are groups of users with similar preferences and the use of the word \emph{community} does not imply closer friendship ties between users.}. {\it Once a summary of the evolving groups has been formed, several other interesting questions can be formed}: Do users form polarized and insular groups? Does one group dominate or drive out other groups? Is there movement between groups?  How can we design online communities to foster cross-group understanding? How do external events affect these dynamics? What are the evolving interest patterns and what is driving them?

We apply this methodology to a social news site, named \emph{Balatarin}\footnote{balatarin.com} (translated \emph{The Highest}), which is a mainly Persian-language website. This platform is suitable for our purpose because it played a significant role during an important political event, the Iranian post-election uprising in 2009, dubbed the Green Movement. Balatarin became a hub for disseminating information and a space for people to exchange opinions, propose ideas and even organize to take action to protest in the real world. Some of the more well-known US-based examples of similar social news sites are Reddit\footnote{www.reddit.com}, Slashdot\footnote{slashdot.org} and Digg\footnote{digg.com}. 


Our methodology: 1) produces a novel visualization of political dynamics throughout the 4-year duration of the data, 2) finds politics-based evolution paths in multi-issue contexts, and 3) extracts user preferences for text and source of content. We are able to observe the patterns at different granularities by producing summaries at multiple scales and at different times. We focus on four example paths and show that as much as 40\% of users stayed in the same path after one year, indicating an implicit yet enduring community of users with consistently similar preference for content. We also find highly specific and persistent themes within some paths, relating to issues (such as international relations) or political orientations (such as pro Green Movement). The visualizations shows that an external event (post-election uprising) had a sudden effect on these dynamics, causing major reorganization of communities. Finally, we evaluate the coherence within each path by studying the entropy of publication sources from representative content and find that recurrence of domains within detected paths doubles, triples, or quadruples compared with articles drawn at random. We find that the paths are not insular and there are merges between them as well as content overlaps. No one group or path becomes so dominant as to drive out others, however, following the election crisis there is a shift in focus and paths reorganize around the Green Movement.   {\em  The authors find it very appealing and  instructive  that such a detailed summary could be reconstructed by employing a  completely unsupervised and automated set of tools} that assumes no knowledge of the underlying events or the background of the users. Presented with such a summary, a decision maker or a researcher can then dig deeper and fill in the relationships and connections with the external events that the group was responding to and participating in. The detailed results and associated commentaries are discussed in section \ref{Section:Results} and demonstrated in Figure \ref{Community Evolutions}.


In the next section we explain the steps of the methodology and include our proposed evaluation method. Section \ref{Section:Experimental Setup} describes the implementation of the methodology on our dataset and details its results. Section \ref{Section:related work} presents an overview of related work and Section \ref{Section:conclusion} discusses further ideas and concludes the paper.   

\section{Description of Methodology}
\label{Section:paths}

In this section we will describe the steps of the methodology: defining the network and implementation of community detection and evolution in successive times. We then produce content summaries of evolving communities and propose a method to evaluate the results.

\subsection{Community Evolution}

\begin{figure}[h]
\begin{center}$
\begin{array}{cc}
\includegraphics[width=1.5in ]{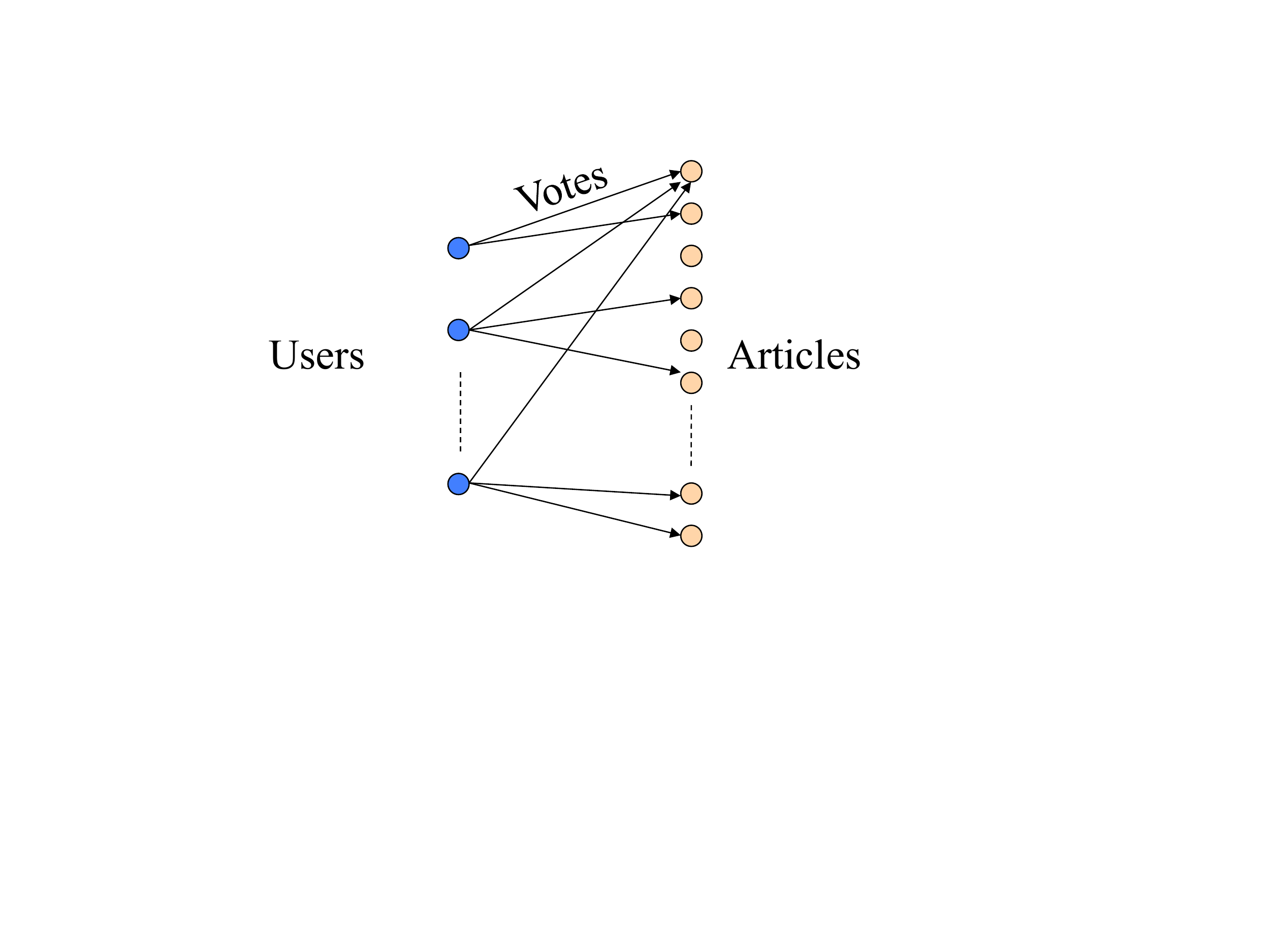}& 
\includegraphics[width=1.35in ]{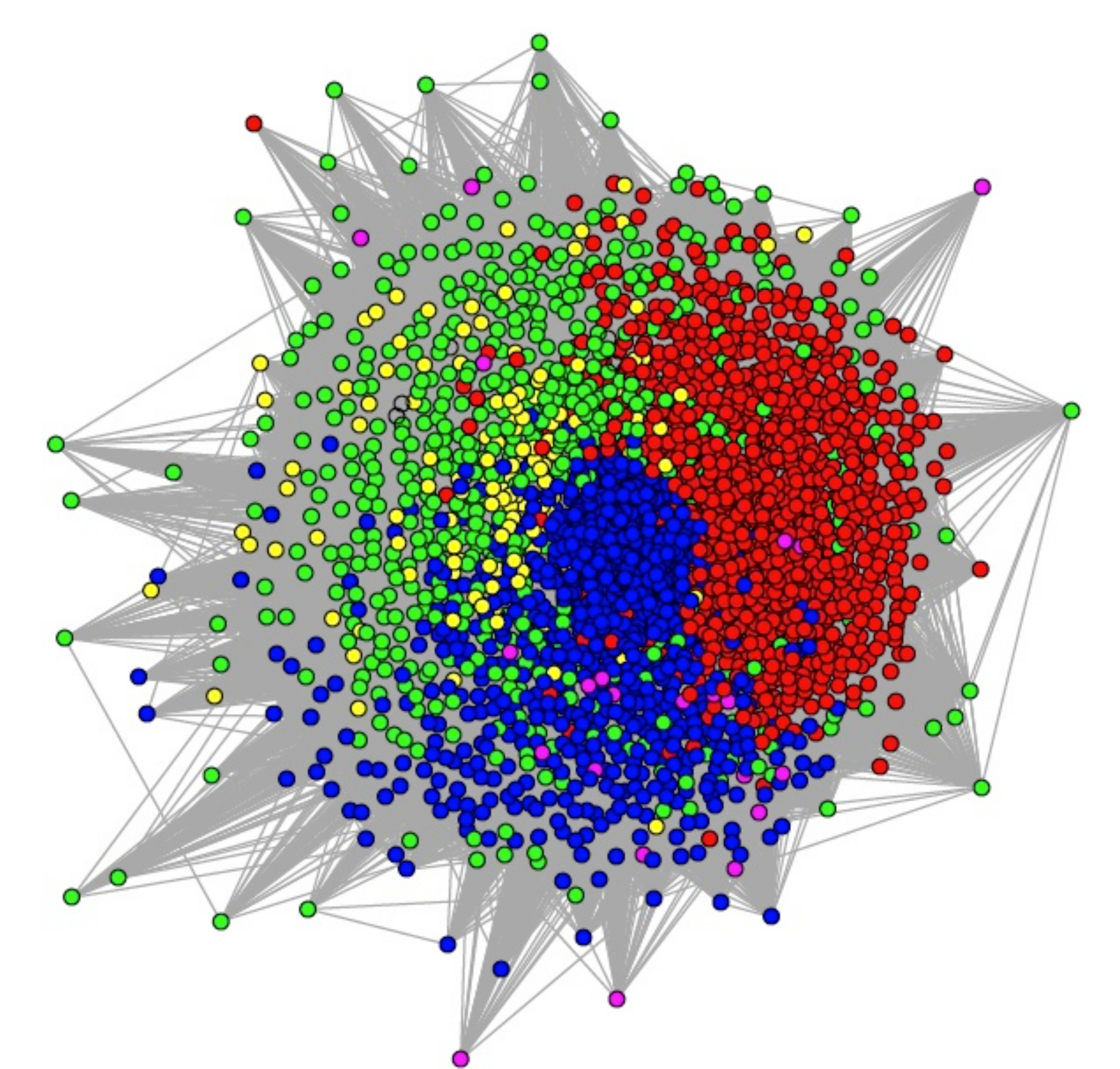}
\end{array}$
\end{center}
\caption{(Left) Bipartite graph of users and articles. (Right) Example of projected graph of users and the communities found in a one month time frame of data, each community in a different color.}
\label{bipartite}
\end{figure}

To group users who vote similarly, we define a bipartite network of users and articles where each edge is a vote cast by a user to an article. Figure \ref{bipartite} illustrates this structure. We project this bipartite network onto a weighted unipartite (single-mode) graph consisting of users only, where the weight of an edge between two users reflects how similarly they vote. The edge weight between a pair of users $(x,y)$ is computed using the Jaccard Index:
\[W_{jaccard}={{n(X \cap Y)}\over {n(X \cup Y)}}\]

 where $X$ and $Y$ are sets of articles voted for by user $x$ and $y$ respectively, and $n$ stands for set cardinality. 
\newline

In the study of network topologies one of the most widely used measures of community formation is the \emph{modularity} metric \cite{Girvan2002}, which compares the number of edges between vertices belonging to the same community to the expected number of edges among the same nodes in a null model-- i.e. a random graph with the same degree sequence. We use the expression for modularity of a weighted graph defined in \cite{PhysRevE.70.056131} as:
\[Q = {{1\over2W}\sum_{i,j}(W_{ij}-{s_{i}s_{j}\over2W})}\delta(C_{i},C_{j})\]
where \(W_{ij}\) is the weight of edges between vertices i and j, \(W\) is the sum of the weights of all edges and \(s_{i}\) is the strength of vertex \(i\) defined as the sum of the weights of edges adjacent to the vertex. \(C_{i}\) is the community that vertex i belongs to and \(\delta\) is the Kronecker delta. The expression \({s_{i}s_{j}\over2W}\) computes the expected number of edges between vertices i and j in the null model.

To find sequences of such vote-based communities, we first construct bipartite graphs and their single-mode projections for the data in consecutive time frames. Then, using a fast modularity maximization algorithm \cite{Clauset2004}, we find communities for each time frame
Figure \ref{bipartite} shows a visual example of communities found in a one month time frame of our dataset that will be described in detail later in the paper. 

For every pair of successive time frames we compute transition probabilities between every community pair $C_i$ and $C_j$ in times $t_1$ and $t_2$ and construct a matrix of transition probabilities. More specifically, each element $P_{ij}$ is computed as: 
\[{ P_{ij} = Pr(x \epsilon C_j (t_2) | x \epsilon C_i(t_1))} \enspace \]

In this matrix, the $C_j(t_2)$ with largest transition probability from $C_i(t_1)$ is the the community in $t_2$ where most of the users in $C_i$ in the previous window move to. Based on highest transition probabilities for every pair of communities in consecutive times, we create a visualization of paths of opinion-based communities.  

\subsection{Representative Content}
\label{Section:RelevantArticles}
The evolving communities detected in the previous section will define the skeleton of voting behavior among users. In order to characterize the nature of detected communities and add a layer of meaning, we first find the articles most preferred by each community. 

\begin{figure}[h]
\centering
\includegraphics[width=2in]{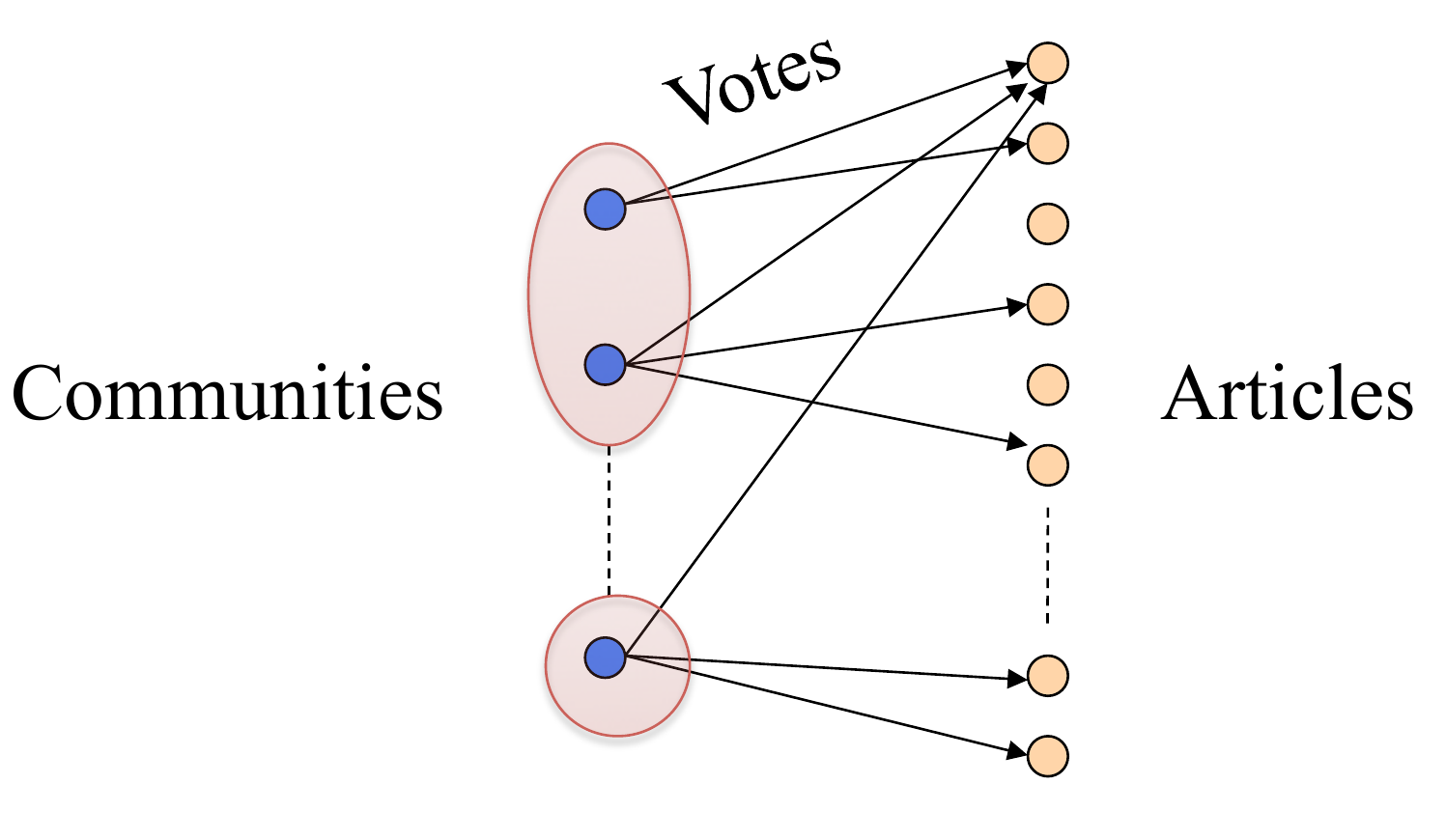} 
\caption{Graph of user communities (shaded ovals) and articles.}
\label{communityBipartite}
\end{figure}

Intuitively, articles preferred by a community will demonstrate a high level of (positive) deviation from the number of votes they are expected to receive from that community. Considering the network of communities and articles with each vote connecting a community to an article (Figure \ref{communityBipartite}), one can construct a random graph such that the degree sequences (i.e. number of votes cast by users in communities and received by articles) are preserved. The random graph is created by connecting an edge coming out of a community to one going into an article uniformly at random. We compute the expected number of edges between communities and articles in this random graph and find the deviation from the true number of edges observed in the data. In the random graph, the expected number of votes given to article $A$ from users in community $C$ will equal: 
\[{E(A,C)={n(C)\over N}n(A)}\]
where $n(C)$ is the total number of votes cast by users in community $C$, $n(A)$ is the total number of votes received by article $A$, and $N$ is the total number of votes cast by all users to all articles.

The deviation score can be computed using the following expression:
\label{Equation:preference}
\[{Score(A,C) = {{(O(A, C)-E(A,C))^2} \over E(A,C)}}\] 
where $O(A, C)$ is the observed number of votes received by article $A$ that come from users in community $C$. We compute this for the cases where  $O(A, C)>E(A,C)$ in order to obtain only those observations that are more popular than expected. This score is inspired by Pearson's $\chi^2$ test statistic \cite{doi:10.1080/14786440009463897}.

Using this expression, we rank articles for every community belonging to a time frame and create a list of most representative articles for each community. We expect that the articles representing each community will have similarities in their content that signify a difference from other communities, and that this preference within each community will carry over through the whole evolution path.


We will now extract a more granular characterization of content reflected in each evolution path. For this purpose, we consider the ranked list of representative articles within each community. Given that each article that is posted to the site includes a title and a summary of its content\footnote{This is almost always the case in online forums and social news sites.}, we use a bag of words model to find the deviation between the words used in representative articles for a community and the rest of the articles posted in a time frame. To compute this deviation, we find term frequencies for all the words in each time frame. We then find term frequencies for the top representative articles per community and normalize term frequencies between 0 and 1. Computing the difference between the two mentioned values provides a deviation score for each term: 

\[{\textrm{Score}_T = {tf_{T,C}(t) \over \max_T{tf_{T,C}(t)}}  -  { tf_T(t) \over \max_T{tf_T(t)}}} \enspace \]

where $tf_{T,C}(t)$ is the term frequency for term $T$ in community $C$ at time $t$ \footnote{Although this process of scoring is similar to term frequency--inverse document frequency (tf-idf) weighting \cite{manning2008introduction}, note that we are not ranking documents and are instead finding a normalized ranking of terms only, so we do not use inverse-document-frequencies.}. We rank the words that belong to each community based on their score. We can regard this ranked list of terms as automatically generated summary tags for each community. Similarly, aggregating top words for each community along its evolution path and finding the most frequent terms over each path will automatically generate summary tags for each evolution path.

Finally, each article includes a URL link to its source of publication. Extracting the domains from URLs of representative articles in each community and aggregating over its complete evolution path provides us with a list of content sources representing each path. 

At this point, we will have a summary visualization of the overall dynamics over time, a set of relevant words and domains (i.e. publication sources) most representative of each evolution path, as well as the capability to drill down to any specific time frame and get a list of representative words and publication sources for each community at that time. Finally, for each community at any time frame, a ranked list of specific representative articles and the url to the full article is available for an in-depth examination. 
\subsection{Evaluation} 
\label{Section:evaluation}

Community detection algorithms have been evaluated on various randomly generated benchmark graphs with community structure (refer to a review paper by Fortunato for a summary of these benchmark graphs \cite{Fortunato2010}). 
Nevertheless, as is the case with our data, typically there is no ground truth available or existent. So in this paper we devise two methods to evaluate communities and their evolution paths. First, we build a simulation model that follows mechanisms of a social news website with reasonable parameter values, and see how well the algorithm finds the ``true" community structure based on (empirically unobserved) individual positions on an opinion space. In other words, we are producing a specific benchmark graph for our dataset which includes a ground truth. Next, we evaluate whether the community evolution paths are meaningful by measuring source entropy within each path. We will now describe these processes in more detail.


In the first method we begin by assigning each user a position on a 2-dimensional Cartesian space that will represent the underlying opinion space\footnote{While for clarity this simulation assumes a 2-dimensional opinion space, we make no such assumptions in the general methodology.}. 
Users are randomly placed according to a normal distribution around one of four equidistant center points in the four quadrants. The position of users is considered the ground truth, with each user belonging to one of the four communities specified by the four quadrant centers. Given this structure, a k-means algorithm that uses the (otherwise unobserved) user positions can find the four user clusters with relative ease thus serves as an approximate lower bound for error in detecting communities. We then generate a set of articles by randomly selecting users who will each post articles and votes. Each generated article is positioned in the opinion space according to a Gaussian distribution near the user who posts it. Each user will vote for an article with some probability, if that article is positioned closer than a certain threshold to him/her in the political space, thus an article is likely to get a vote if it's close to a reader's opinion. The result of this process is a set of users, articles, and votes which we then use as a simulated graph for a social news platforms. 
Complete details of the simulation parameters and more detail on results are available on the website.  
\footnote{
http://rostam.ee.ucla.edu/mediawiki/index.php/Social\_
News\_Simulation}

\begin{figure}[h]
\centering
\includegraphics[width=3in]{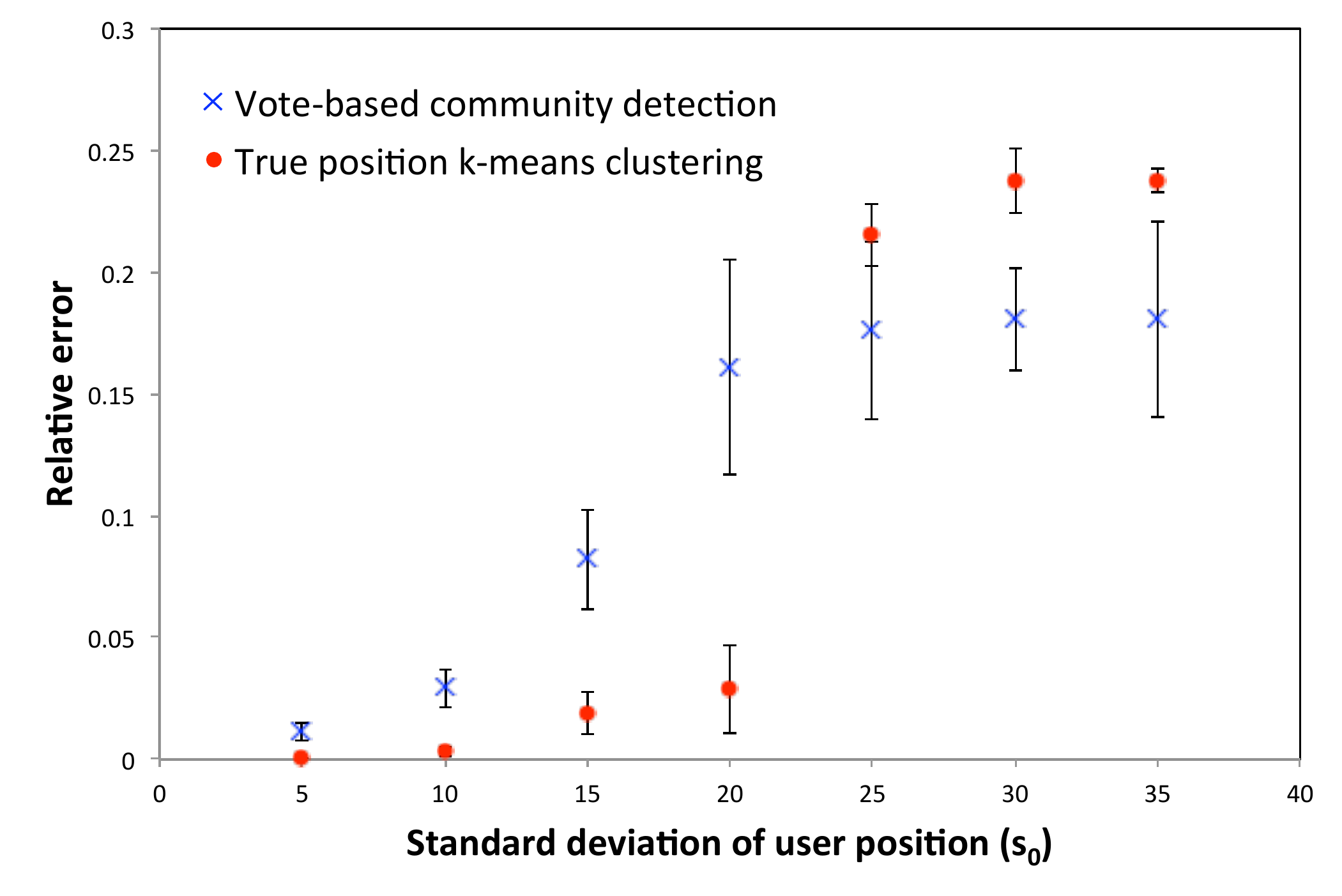}
\caption{Relative error vs. standard deviation of user positions in the opinion space. The jump in the k-means error is due to the fact that true user memberships are no longer recognizable. Relative error is computed based on pairs of users that are classified incorrectly together or separate.  500 users were generated. Results are based on an average of 10 simulations. Error bars mark two standard deviations.}
\label{evaluation}
\end{figure}

We simulate this data with different variances for the aforementioned Gaussian distributions. We then run our network-based community detection algorithm on this graph and compute relative error as we change the variance of underlying data generation process (simulation model). 
Figure \ref{evaluation} compares the results of community detection (based on votes) with k-means clustering (based on true positions of users) as the standard deviation of the Gaussian distribution used to generate user positions changes. The algorithm is generally robust and successful in finding true underlying clusters while error increases with the standard deviation of user positions (i.e. as users are more scattered). When the value of standard deviation reaches the mid-point between the two centers, neither k-means nor the network based algorithm can detect clusters correctly. This simply means that users are distributed such that clear clusters do not exist anymore. 
In addition to the above simulation-based evaluation, we propose an indirect way of evaluating the full evolution path for each community. This method is based on finding whether throughout the length of a community's evolution path, there is a preference for a few sources of publication. Since the votes are cast to completely different articles, there should be no expectation that their sources be the same unless users in each evolution path are favoring certain sources of information over others, an indication of common underlying preferences. 

We aggregate the top $n$ representative articles over all the time frames in a community evolution path. We then calculate the \emph{Shannon Entropy} \cite{shannon1949mathematical} of the source of these articles (as indicated by their domains). This will signify the amount of source variation over top preferred articles for each evolution path: 

\[{\textrm{Entropy}(C) = -\sum_{i} {p_i log_2(p_i)}}\]
\label{Equation:Entropy}

where $p_i$ is the probability that an article from source $i$ is in the top $n$ most preferred articles of community $C$. 

A lower entropy value indicates lower variation and higher uniformity in sources of articles. Entropies found for evolving communities are then compared to entropies from sets of articles drawn at random. We generate the random sets by randomly choosing votes, finding which articles the votes were cast for, and then extracting the domain of the article. We randomly choose \emph{votes} rather than randomly choosing \emph{articles} because we want the articles with higher votes to have a higher probability of being chosen. This is important because the list of most preferred articles in each community is also based on the preference of a community's users to vote for that article. 

We then compute the \emph{effective number of sources} in an evolution path as $2^{Entropy}$ and compare with that of the randomly selected sets\footnote{This measure is used in Ecology as the \emph{effective number of species}\cite{Hill1937} in an ecosystem. Another metric for diversity is found by comparing the effective number of sources with the number of unique sources in each set. Using this metric we reached similar results.} and compute the ratio as: 
\[\textrm{Relative Recurrence}= {2^{Entropy(random)} \over 2^{Entropy(path)}  } \]

A higher recurrence in sources of information compared with the randomly drawn dataset will strongly suggest that the evolution paths are highly preferential toward certain sources, corroborating that they are meaningful. In the next sections we will demonstrate this methodology on a real dataset, where the above explanations will become more clear with example.


\section{Experimental Setup}
\label{Section:Experimental Setup}
\subsection{Data Description}
We apply our methodology to a social news website with article link submission, voting, and commenting systems. The website, named \emph{Balatarin} (translated \emph{The Highest}), is a mainly Persian-language social news site that played an important role during the Iranian post-election protests in 2009, dubbed the Green Movement. The website became a hub for disseminating information, as well as a space for people to exchange opinions, propose ideas and even organize to take action to protest in the real world. The massive uprising marked a turning point in Iranian politics and while a great deal of media attention\footnote{As an example, read Washington Post article titled ``Twitter Is a Player In Iran's Drama", published June 17, 2009} was paid to the role of Twitter in the protests, less consideration was given to Balatarin, mainly due to the language barrier. Nevertheless, inside Iran and within the Persian-speaking population Balatarin was among the most prominent social web entities at the time. Balatarin is similar to Reddit in that  it does not have an explicitly defined friendship network, yet similar to Digg in that it focuses on positive votes to rank articles.

\begin{figure}[h]
\centering
\includegraphics[width=2.5in]{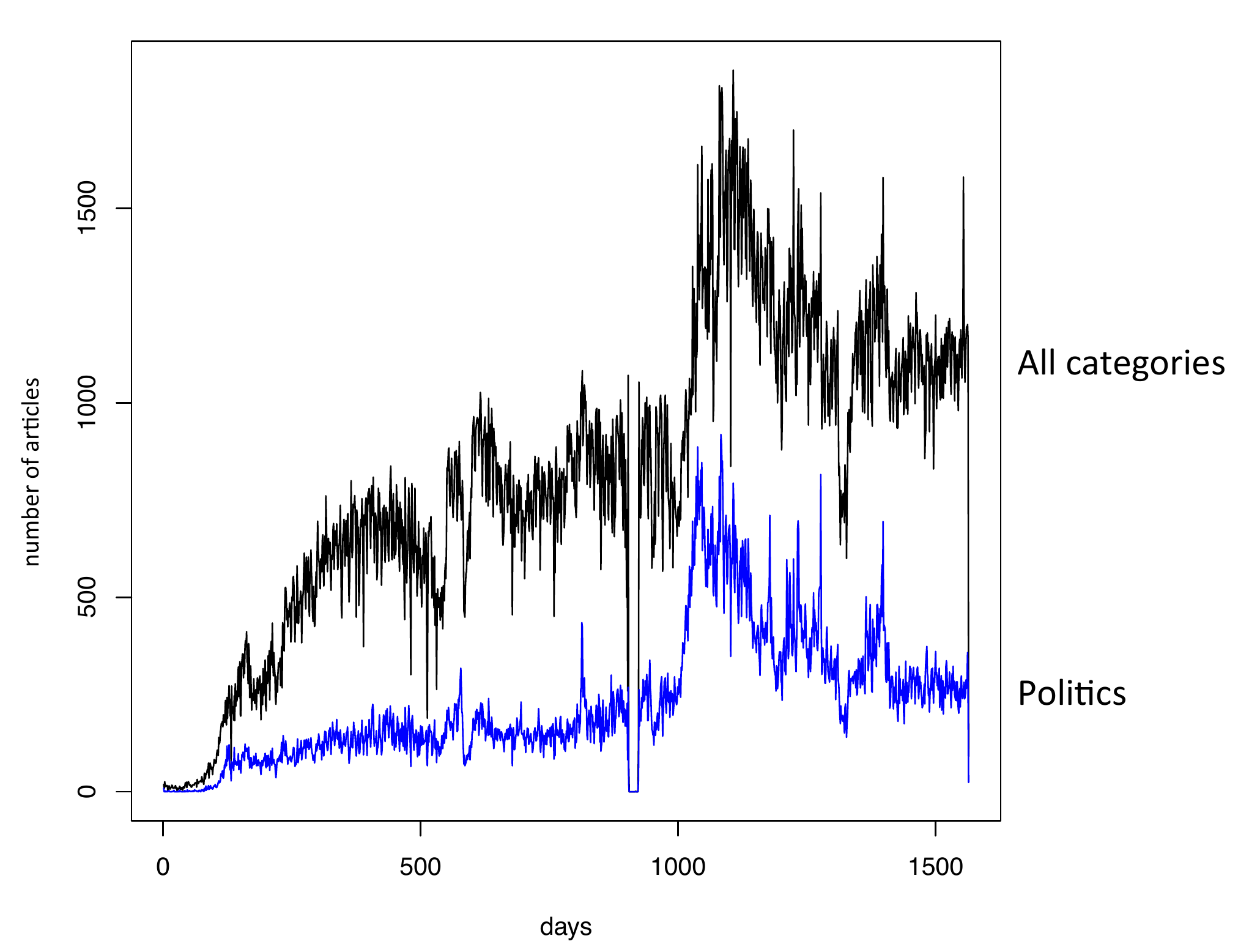}
\caption{Timeline of articles posted to the site (top graph is all articles, bottom is only articles in the politics category). The data spans over 1500 days.}
\label{linksTimeline}
\end{figure}


\begin{figure*}[htbp]
\centering
\includegraphics[width=7in]{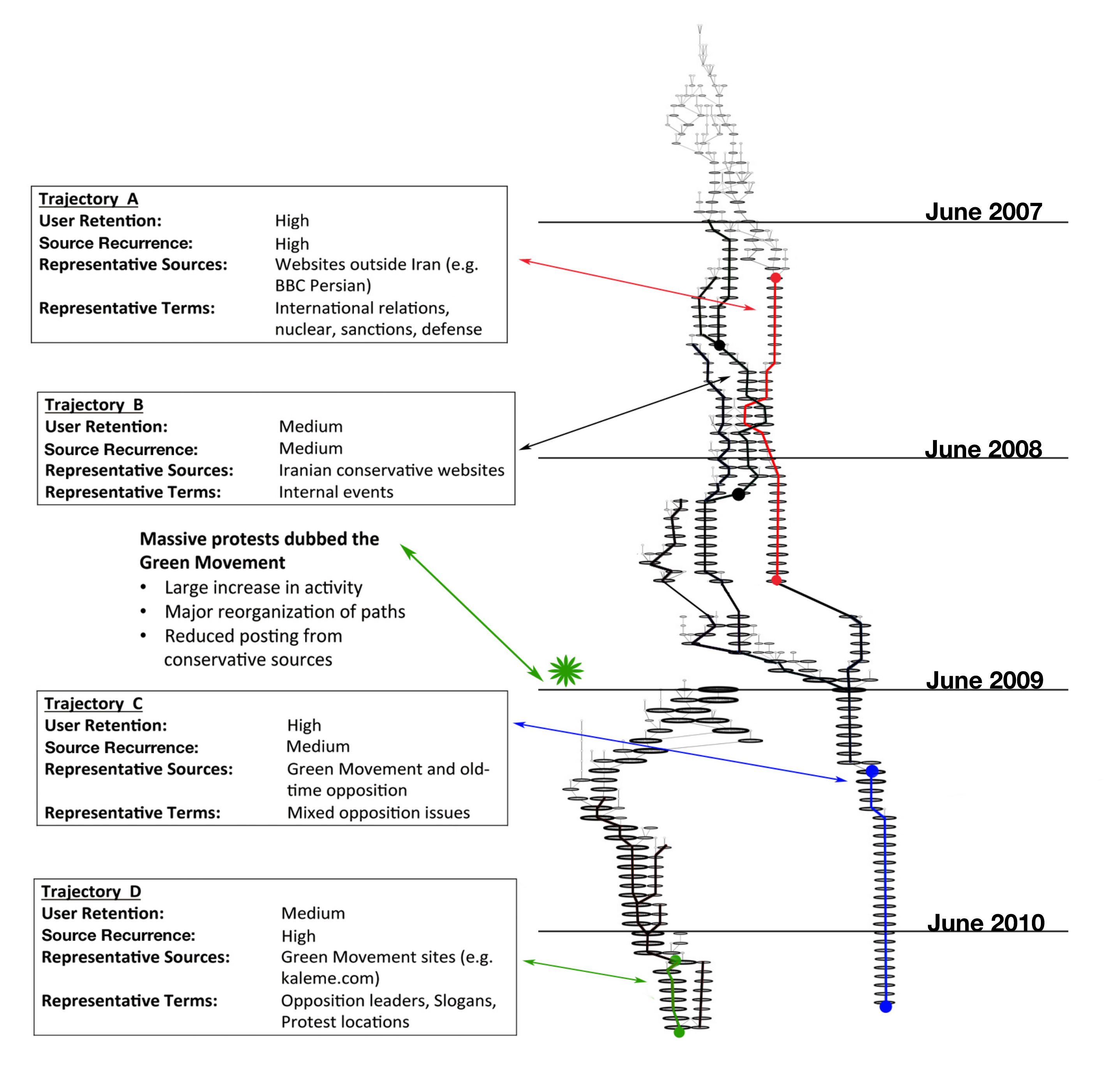}
\caption{Paths illustrating evolving communities in \emph{Balatarin.com}. Time begins on top of the figure and progresses downward, oval shapes represent communities and their sizes correspond to community size (the horizontal position of the communities is merely chosen for ease of visualization). Boxes summarize characteristics of four example paths labeled as A, B, C, and D and delineated by large dots on the graph. A significant event (Iranian post-election uprising in June 2009) marks a transition in the dynamics of the site. Note that there are several other paths that are observable in this graph, and while we have only chosen four as demonstration, other paths are of similar quality to the selected paths.} 
\label{Community Evolutions}
\end{figure*}

The dataset includes a total of over 1.2 million articles, 26,000 users and 31 million votes posted from August 2006 to November 2010. Less than 3\% of users are responsible for more than 55\% of the votes. The articles are tagged according to their category and we will focus our attention on the articles in the Politics category (a total of 352,000 articles) since finding trends within non-related content categories will not be a meaningful or desirable task. Figure \ref{linksTimeline} shows  a timeline of number of articles posted to Balatarin as well as the number of articles in Politics. The sudden rise in the number of articles coincides with the 2009 protests\footnote{The short sharp drop to zero marks a shut-down due to an attack on the site in February 2009.}. 


To investigate the data over time, we choose a 30-day time frame and slide this frame over the duration of the data to form temporally consecutive datasets (users, articles, and votes in each time frame). Sliding the frame two weeks at a time produces 110 time-frames over the whole duration of the data. 

\subsection{Results}
\label{Section:Results}
Figure \ref{Community Evolutions} shows evolving communities over 110 overlapping time frames, starting at the launch of the website in 2006 on top of the figure and progressing downward. Each oval shape represents a community and  communities placed on the same row belong to the same time window \footnote{Graph was generated using the PyGraphviz library in python.}. Size of the ovals reflects the number of users in the community (community sizes range from 10 users to over 3000 users) and communities in consecutive times are connected as described in the previous section. 

Distinct evolution paths of different durations can be observed and events such as birth, death, merge, split, growth and contraction of communities are evident along the paths. Furthermore, the effects of the Iranian post-election protests in June 2009 is readily evident as a sudden increase in community sizes occurs at the onset of the event. This is in agreement with the increase in number of articles (Figure \ref{linksTimeline}) which almost doubles during this time. In addition, there is a shuffling of paths and there are sizable merges and re-formation of paths after the event.  Thus, similar to its effects in the real world, this event has had a significant impact on the dynamics of the user population on the site. We choose four paths (labeled A,B,C, and D) to investigate further in the next sections. These were chosen such that we have a number of paths occurring at different times and not due to any superiority of quality; other paths are of similar quality to the selected paths. These paths are marked on the figure, two of them corresponding to a time prior to the June 2009 event, and two of them belonging to a time after the event. Following the steps in Section \ref{Section:RelevantArticles}, we produce representative terms and domains for each path. Table \ref{Table:content} lists these results. At this stage, we can step into a finer granularity by focusing on specific points that may be of interest, such as a merge between two communities. Specific terms, domains, article summaries, and urls representative of each community are readily available for further investigation through simple queries. 


\begin{table}[h]\scriptsize
\caption{Summary of domains and terms associated with four example evolution paths. Terms have been translated from Persian to English.}
\label{Table:content}
\centering
\setlength{\tabcolsep}{6pt}
\begin{tabular}{p{0.5cm} p{3cm}p{3cm} }
\hline\noalign{\smallskip}\hline\noalign{\smallskip}
Path & Domains  & Terms \\
\noalign{\smallskip}
\hline
\noalign{\smallskip}
 A & 
www.bbc.co.uk	
www.dw-world.de
www.roozonline.com
www.isna.ir	
www.noandish.com
&
Nuclear,
America,
Iran,
People,
Republic,
Russia,
Iraq,
Israel
 \\
\noalign{\smallskip}
\hline
\noalign{\smallskip}
 B & 
www.alef.ir 
www.youtube.com
www.noandish.com
www.tabnak.ir
 
farhadheyrani.blogspot.com
&
Photo, 
Leader, 
Torture,
Mortazavi,
Prison,

Father, 
Child,
Public,

Ahmadinejad

\\
\noalign{\smallskip}
\hline
\noalign{\smallskip}
 C & 
www.rahesabz.net
zamaaneh.com	
www.radiofarda.com
www.dw-world.de
news.gooya.com
&
Prison, 
Participation, 
Rights,
Karroubi, 
Government, 
Country,
Political,
Arrest
\\
\noalign{\smallskip}
\hline
\noalign{\smallskip}
 D & 
www.youtube.com
www.kaleme.com
www.rahesabz.net

iarandoost657.blogspot.com
gomnamian.blogspot.com
&
MirHossein, 

Allah-o-Akbar, 

Mousavi, 
Islamic, 

Slogan, 
Square, 
Security,
Mehdi [Karroubi]
\\
\noalign{\smallskip}
\hline
\noalign{\smallskip}\hline\noalign{\smallskip}
\end{tabular}
\end{table}

\subsection{Evaluation}
\label{Section:experiment evaluation}

Manual inspection shows evidence of similarity of preference between users in each community both in text and in sources of content. In some extreme cases small communities demonstrate strong preference for certain websites to the point where the links most associated with that community all belonged to the same domain. An example is a community of 18 users in January 2009 whose top 10 preferred articles all came from the pro-government website (www.fararu.com) and had very high graph density (all users in the community voted exactly the same way). This occasional extreme uniformity in source of articles (as evident by the domain) hints to a possibly organized off-site effort by a group of users somehow affiliated with the website, paid by the same entity, or dedicated to advocating a cause. 

\subsubsection{User Retention}

Since our goal is to group users solely based on their vote similarity, we do not define and utilize \emph{core} users to find evolving communities (a few papers have proposed finding communities based on core users \cite{Seifi:2012:CCE:2187980.2188258} \cite{Wang:2008:CCA:1428392.1428416}). Therefore, because evolution paths are inferred by computing transition probabilities between consecutive time frames, it is not clear whether the paths will remain meaningful and consistent after several time steps. If at each step a number of users leave and new users join the community, will any of the same users remain after several time frames? Will there still be content coherence within the whole path? Will it be reasonable to assume this is the same evolving community after so many time steps? To answer these questions we compute \emph{user retention} by studying membership within a path across several time steps.  For a path $P$, we compute user retention after $\Delta \tau$ periods from time $\tau_i$ as:

\[\textrm{Retention}(P,\Delta \tau) = { {n(P(\tau_i) \cap P(\tau_i+\Delta \tau))}\over {n(P(\tau_i))}} \enspace \]

where $P(\tau_i)$ is the set of users in path $P$ at time $\tau_i$. 

\begin{figure}[h]
\centering 
\includegraphics[width=3in]{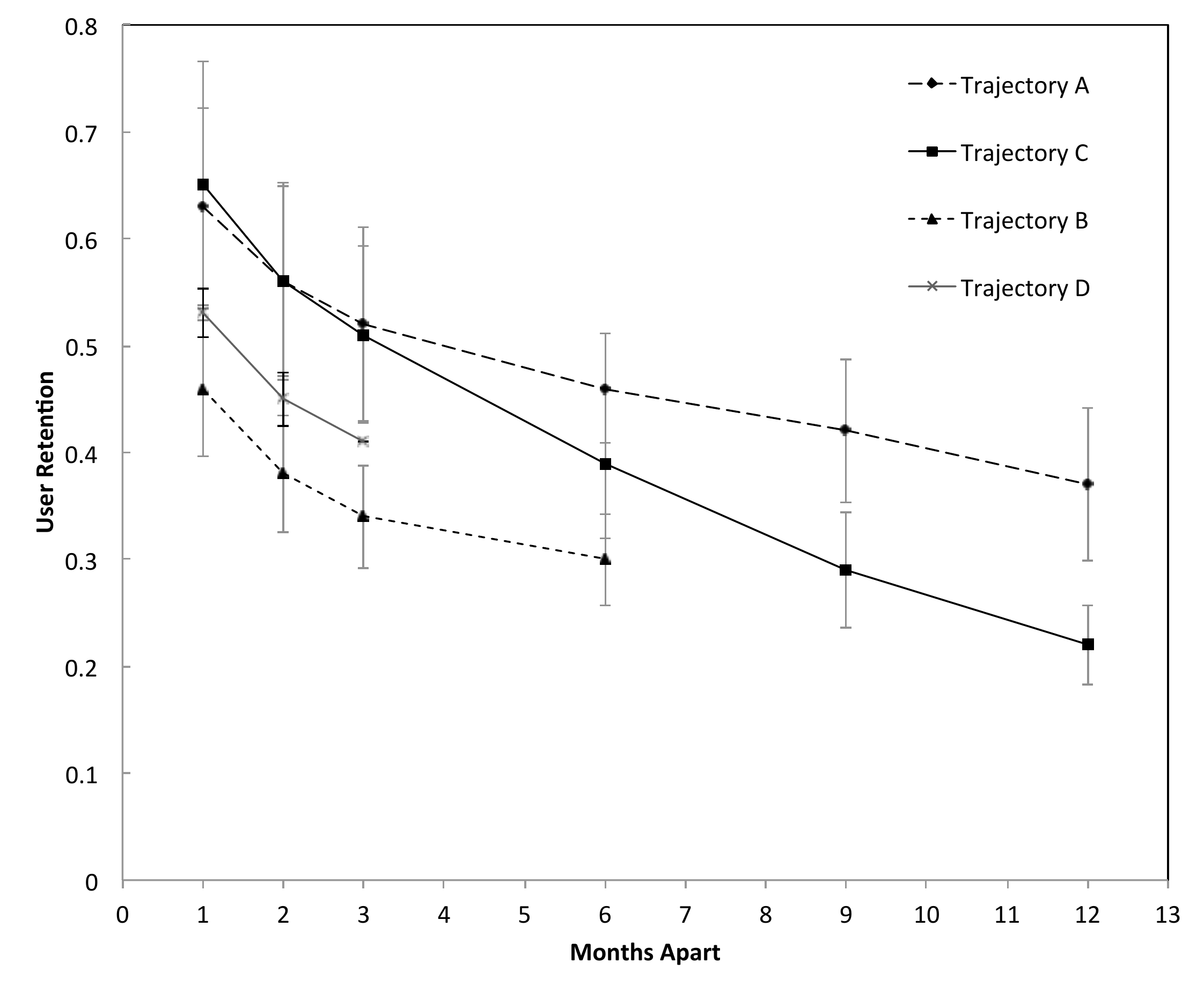} 
\caption{User retention (average fraction of users remaining in path) vs. $\Delta \tau$ for community evolution paths A, B, C, D}
\label{userRetention}
\end{figure}

Figure \ref{userRetention} illustrates the average retention for different $\Delta \tau$ values (retention is averaged for $\tau_i$'s spanning the whole path) for four different evolution paths. Note that two of the chosen paths are longer (spanning more than one year) and two are shorter. The results show that evolution paths have reasonably high user retention. Paths A and B have more than 50\% user retention within 3 months, and between 20\%-40\% user retention after 1 year (24 evolutionary cycles in on our algorithm). In the worst case, 20\% of the users are voting similarly after one year which is significant considering the fact that these communities are not based on any explicit connections and that within a year there is natural turn-around in a site's user population.

\begin{table}[h]
\centering
\caption{Relative recurrence of domains in each path, demonstrating that paths are highly preferential toward certain content sources.
}
\begin{tabular}{|c|c|c|c|c|} \hline
Path & A& B& C& D\\ \hline \hline
Relative recurrence & 3.60 & 1.74 & 2.31 & 2.41 \\ \hline
\end{tabular}
\label{Table:recurrence}
\end{table}

\subsubsection{Source Recurrence}
Table \ref{Table:recurrence} lists relative recurrence of sources within each of the four selected paths A,B,C, and D as explained in Section \ref{Section:evaluation}. We observe that all four evolution paths have an increase in recurrence of information sources. We see as much as 3.6 times more recurrence of sources compared to the set drawn randomly (proportionally to an article's votes), demonstrating strong preferences toward some sources of information.

\section{Discussion of Results}
\label{Section: Interpretation}

We will now combine all the information that can be gleaned from the paths and their summaries as produced through our automated and unsupervised process. We focus on the selected paths A,B,C, and D as labeled on Figure \ref{Community Evolutions} and summarized in Table \ref{Table:content}. Note that there are several other paths and we have only chosen to focus on these four as examples. We can see that the massive protests, dubbed the Green Movement, that took place after the Iranian presidential elections on June 12th 2009 create a significant disruption in the organization of paths on Balatarin. Paths A and B are prior to this disruptive event, whereas C and D occur after this event. Also, paths A and B overlap in time and path D overlaps with the end of path C, therefore we can compare and contrast them respectively. 

\begin{itemize}
\item \textbf{Path A:} 
Table \ref{Table:content} shows that this path is formed mainly around the issues of Iran's international relations, including the nuclear issue, as well as relations with the US, Russia and Europe, Israeli-Palestinian conflict, and the Iraq war (terms related to sanctions and defense are also among the top terms but are not listed in the table due to space limitations). This path favors articles from prominent news agencies outside Iran (which the Iranian authorities generally do not approve of) such as the BBC Persian\footnote{www.bbc.co.uk} and the Germany-based Deutsche Welle\footnote{www.dw-world.de} but also articles from some news agencies within Iran such as that of Iranian Students' News Agency \footnote{www.isna.ir} which at the time published content close to reformist groups. This path has high user retention and high recurrence of its sources of content. 

\item \textbf{Path B:} 
This path is focused more on Iranian internal issues and while it does not demonstrate strong loyalty to specific domains, it favors articles from conservative websites inside Iran (www.alef.ir and www.tabnak.ir both belong to conservative Iranian statesmen). This path shows more variability in sources of content and its top domains include websites belonging to reformists (rivals to the conservatives) as well. Path B shows the lowest user retention among all four example paths. We observe that the paths before the election were more issue based, focusing on international versus internal issues.

\item \textbf{The Green Movement:} This major external event occurred in June 2009 when the Iranian government violently crushed large protests. We observe a significant reorganization in paths and their contents. The presence of government news sources and conservative Iranian websites (such as those in path B) has almost vanished in major paths. While these websites do appear in smaller intermittent communities, the communities fail to continuously stay active and create a path. We found that a number of users from Path B were absorbed into other paths (possibly because of a change in their political position). 

\item \textbf{Path C:} This is a long-lasting path that favors news and analysis from news agencies outside Iran as well as sites affiliated with the Green Movement (e.g. www.rahesabz.net). Although very much focused on the aftermath of the Green Movement, this path demonstrates more diversity in its sources of content. While this path is of similar length to path A, its user retention drops faster than that of path A after one year.

\item \textbf{Path D:} This path demonstrates clear political leanings through its very high recurrence of content sources that are well-known websites affiliated with the Green Movement (www.kaleme.com and www.rahesabz.net) and its text tends toward names of Green Movement leaders, slogans, and protest locations as evident in Table \ref{Table:content}. Although of similar political orientation, a difference between paths C and D is that path C is more focused on news and analysis from established news agencies whereas path D has a preference for blogs and youtube videos, and in terms of content it focuses on eyewitness accounts and protest organization. Some of these weblogs seem to have been created solely for reporting specific protests and may have few posts, some others have been shut down. 



\end{itemize}

In addition to the above summaries, because domain and term rankings are created for each community at a time frame, the results provide a multi-scale capability where we can in fact focus further on summaries of a path at a certain time frame and compare dominant themes across different times as needed. 

\section{Related Work}
\label{Section:related work}

Clustering and community detection methods are in essence network summarization tools  \cite{Girvan2002}\cite{Clauset2004} \cite{Blondel}\cite{PhysRevE.76.036102}\cite{PhysRevE.76.066102}. A survey paper by \cite{Fortunato2010} provides a comprehensive summary of this field. Building on this literature, a growing body of work has been produced on community evolution, varying from works on evolutionary clustering \cite{Palla2007}\cite{Chakrabarti:2006:EC:1150402.1150467}\cite{Wu:2009:CEC:1651437.1651444} and communtiy detection in dynamic social networks \cite{Tantipathananandh:2007:FCI:1281192.1281269} to processes that also optimize for smoothness in temporal evolution \cite{Lin:2008:FFA:1367497.1367590} or use community cores in evolving networks \cite{Seifi:2012:CCE:2187980.2188258}. A categorization and review of community evolution methods is presented by \cite{6152081}. These works focus solely on the network structure. 

A number of recent papers focus on using content alone to create summaries of text, such as opinions \cite{Ganesan:2012:MGU:2187836.2187954}, product reviews \cite{Liu:2005:OOA:1060745.1060797}, political leanings \cite{Fang:2012:MCO:2124295.2124306}\cite{Kaschesky:2011:OMS:2037556.2037607}\cite{DBLP:conf/sigir/JiangA08}, and news streams - more specifically, \cite{Shahaf:2012:TTG:2187836.2187957} create structured summaries of content in the form of narrative maps and \cite{Ahmed:2011:UAS:1963405.1963445} produce story-lines of streaming news. The bulk of literature in this field uses text-based techniques such as language models used in sentiment and subjectivity analyses and topic modeling and are not concerned with user networks. On the other hand, incorporating both the network graph and content, \cite{Jo:2011:WTD:1963405.1963444}  use the citation network between documents to get a better summarization of document content over time (topic evolution), \cite{Lin:2010:PSM:1835804.1835922} track popular events in the social web, and \cite{lin2009jam} summarize activity over time. Yet none of the mentioned papers produce a comprehensive multi-scale map of group behavior among users. 

There is considerable debate whether new online spaces promote diversity or through winner-take-all dynamics exacerbate polarization and conflict in society. While some have hailed the promise of democratic effects of the Internet, others have argued against this notion, asserting that such web-based platforms increase interaction among like-minded people and reduce contact among people of different opinions, leading to fragmentation in society (see for example, \cite{Rowman2002} \cite{Westen98} and \cite{hindman2009myth}). \cite{Adamic:2005:PBU:1134271.1134277} demonstrate political polarization in linking patterns between blogs labeled as liberal or conservative. \cite{Alstyne05}\cite{Rahmandad2011} and \cite{Marvel2011} propose and simulate models of polarization dynamics in populations, Zhou et al \cite{ICWSM112782} jointly classify Digg users and news articles in one of two classes (liberal or conservative) using label propagation starting from a small number of labeled users and articles. Finally, a seminal work in political science \cite{Poole1985} models polarization in American politics through analyzing roll-call votes by members of congress (also see \cite{koford1989dimensions} on dimensionality of these votes).


\section{Concluding Remarks}
\label{Section:conclusion}
Motivated by the challenge of understanding group behavior of user populations in large disorderly data, we devised a novel summarization methodology that produces a multi-scale map of community evolution. The proposed method is fully automated and unsupervised and can be widely applied to other contexts. We used indicators of user preference for content (such as ``likes" or ``votes") and demonstrated that they are a meaningful measure for finding communities in multi-issue contexts. The methodology generates profiles of evolving communities based on their representative content, and evaluates them by measuring recurrence of sources of information preferred by their users. 

Evolution paths found for the real-world dataset in this paper showed high user retention and in varying degrees favored different text and sources of content. We observed that recurrence of sources in articles representing an evolving community at times quadruples as compared with a randomly drawn set of articles, corroborating the reliability of the detected paths. Last but not least, the methodology provides a means to observe data at different granularities by producing summaries throughout the evolution path as well as within each community in one time frame, allowing expert investigators to formulate further inquiries. 

\section{Acknowledgments}
We thank Mehdi Yahyanejad for providing us with Balatarin data and for sharing his valuable insights about group dynamics on Balatarin. We thank Allen Huang for his programming work. This material is based upon work partially supported by the National Science Foundation under Grant No. 1027413. Any opinions, findings, and conclusions or recommendations expressed in this material are those of the authors and do not necessarily reflect the views of the National Science Foundation.

\fontsize{9pt}{10pt}
\selectfont
\bibliographystyle{aaai}
\bibliography{BandariRahmandadRoychowdhury2013}

\end{document}